\documentclass[12pt]{article}

\def\fun#1#2{\lower3.6pt\vbox{\baselineskip0pt\lineskip.9pt
\ialign{$\mathsurround=0pt#1\hfil##\hfil$\crcr#2\crcr\sim\crcr}}}

\title{Photons and static gravity}
\author{L.B.Okun \thanks{E-mail: okun@heron.itep.ru} \\
~~~ \\
ITEP, Moscow, 117218, Russia}
\date{}

\begin{document}
\maketitle

\begin{abstract}

The influence of static gravitational field on frequency, wave-length and
velocity of photons and on the energy levels of atoms and nuclei is
considered in the most elementary way. The interconnection between these
phenomena is stressed.

\end{abstract}

\section{Introduction}

The behaviour of light in a static gravitational field has a long history.
It has been discussed in many monographs on General
Relativity and in vast popular literature. (An extensive list of references
may be found in the recent publications \cite{1,2}.)

In this note we discuss from a single point of view four effects which in the
literature are usually discussed separately:

1) the gravitational redshift of photons flying away from a gravitating body,

2) the increase of energy difference between levels in atoms with the
increase of their distance from gravitating body,

3) the retardation of radar echo from planets,

4) the deflection of star light observed during a solar eclipse.

The relation between the effects 1) and 2) was thoroughly discussed in refs.
\cite{1,2}. This note may be considered as a continuation of those articles.

\section{Redshift and clocks}

The phenomenon of gravitational redshift is widely explained by using
(explicitly or implicitly) a presumed analogy between a photon and a stone:
when moving away from a gravitating body (sun or earth) both a photon and a
stone are supposed to lose energy overcoming the gravitational
attraction.

This explanation, as was stressed recently \cite{1,2}, is wrong. The energy
of a photon $E$ and hence its frequency $\omega = E/\hbar$ do not depend on
the distance from the gravitating body, because in the static case the
gravitational potential does not depend on the time coordinate $t$.
The reader who is not satisfied with this argument may look at Maxwell's
equations as given e.g. in section 5.2 of ref. \cite{3}. These equations with
time independent metric have solutions with frequencies equal to those of the
emitter.

In what follows we will often use both the quantum notion of photon and the
classical notions of electromagnetic wave or even ray of light. To a certain
extent this is caused by the physical nature of experiments in which the
effects we are discussing were observed. For instance, in the famous
experiments by Pound et al. the $\gamma$-quanta were flying vertically in a
tower. The language of quantum physics is necessary to discuss the behavior
of nuclear and atomic levels, and hence of atomic clocks. On the other hand
in the discussion of the radar echo and deflection of light quantum effects
are absolutely not essential.

The proper explanation of gravitational redshift lies in the behavior of
clocks (atoms, nuclei). The rest energy $E_0$ of any massive object increases
with increase of the distance from the gravitating body because of the
increase of the potential $\phi$:
\begin{equation}
E_0 =mc^2(1+\phi/c^2) \;\; .
\label{1}
\end{equation}
The gravitational potential is
\begin{equation}
\phi =-GM/r \;\; ,
\label{2}
\end{equation}
where $G = 6.67 \cdot 10^{-8} \; {\rm cm}^3/{\rm g}\cdot {\rm s}^2$
is the Newton constant,
$M$ is the mass of the body, and $r$ -- the distance from its
center; $M = 2\cdot 10^{33}$ g for the sun, while for the earth $M =
6\cdot
10^{27}$ g.

The universal increase of rest energy of an atom in an excited
and the ground state means that the energy difference between levels also
universally increases by the factor $(1+\phi/c^2)$. As a result of this
increase the energy of a photon emitted in a transition of an atom downstairs
is not enough to excite a reverse transition upstairs. For the observer
upstairs this looks like a redshift of the photon.

A ``semi-competent" observer who knows nothing about the dependence of the
rate of the standard clocks (atoms, nuclei) on the value of gravitational
potential could conclude that photon is redshifted not relatively, but
absolutely. A competent observer knows about special experiments on airplanes
which proved that, in accord with General Relativity, (atomic) clocks run
faster high above the earth (see e.g. refs. \cite{1,2}). Therefore for a
competent observer the apparent redshift of the photon is a result of the
blueshift of the clock.

A naive (but obviously wrong!) way to derive the formula for the redshift is
to ascribe to the photon with energy $E$ a mass $m_{\gamma} =
E/c^2$ and to apply to the photon a non-relativistic formula $\Delta E =
-m_{\gamma} \Delta\phi$ treating it like a stone. Then the relative shift
of photon energy is $\Delta E/E =-\Delta\phi/c^2$, which
coincides with the correct result. But this coincidence cannot justify the
absolutely thoughtless application of a nonrelativistic formula to an
ultrarelativistic object.

It is interesting that one can still speak about the redshift of the photon
if one considers not its frequency, but its wavelength or momentum
\footnote{It is instructive to compare the behavior of $\lambda$ in the
spherically symmetric case with that in a linear potential near the earth's
surface considered in detail in refs. \cite{1,2}.}.  To see this let us
consider the condition that photon is massless:

\begin{equation}
g^{ij} p_i p_j =0 \;\; ,
\label{3}
\end{equation}
where $g^{ij}$, $i, j=0,1,2,3$ are contravariant components of the metric
tensor. In a spherically symmetric potential when photon moves along a radius
we have
\begin{equation}
g^{00}p_0 p_0 -g^{rr}p_r p_r =0 \;\;.
\label{4}
\end{equation}
In the case of standard Schwarzschild metric \cite{3,4,5}:
\begin{equation}
g^{00} =(1-\frac{r_g}{r})^{-1}  \;\; ,
\label{5}
\end{equation}
\begin{equation}
g^{rr} =(1-\frac{r_g}{r}) \;\; ,
\label{6}
\end{equation}
where
\begin{equation}
r_g = 2MG/c^2 \;\;.
\label{7}
\end{equation}

For the sun $r_g = 3$ km.

We identify the covariant vector $p_i$ (and not contravariant vector $p^i$)
with energy - momentum four-vector because
\begin{equation}
p_i =\partial S/\partial x^i \;\; ,
\label{8}
\end{equation}
where $S$ is action. This guarantees that if time is uniform ($g^{ij}$ is
time independent), then energy of a particle is conserved.

As the energy of the photon $E=c p_0$ does not depend on $r$, we immediately
see from Eq.(\ref{4}) that its momentum $p_r$ does depend:
\begin{equation}
p_r =\frac{E}{c}(1-\frac{r_g}{r})^{-1} \;\; .
\label{9}
\end{equation}
The closer the photon to the sun, the larger its momentum.
Hence its wavelength is blueshifted:
\begin{equation}
\lambda = \frac{2\pi\hbar}{p_r} = \frac{2\pi \hbar c}{E} (1-\frac{r_g}{r})
\label{10}
\end{equation}
In that sense (but only in that!) the widely used words about redshift
and blueshift of the photon are correct.

\section{Photon velocity}

With $\omega$ being constant and $\lambda$ decreasing with radius the
velocity of the photon
\begin{equation}
v =\frac{\lambda \omega}{2\pi} = c(1-\frac{r_g}{r})
\label{11}
\end{equation}
also decreases with decreasing radius $r$. In that respect photon drastically
differs from a non-relativistic stone, falling into the sun. (The velocity of
the stone obviously increases with the decrease of $r$.)

The same conclusion could be reached by an explicit calculation of velocity
from the expression for the interval:
\begin{equation}
ds^2 =g_{00}c^2 dt^2 - g_{rr}dr^2 =0 \;\; ,
\label{12}
\end{equation}
where
\begin{equation}
g_{00} =1/g^{00} = (1-r_g/r)
\label{13}
\end{equation}
\begin{equation}
g_{rr} =1/g^{rr} = (1-r_g/r)^{-1}
\label{14}
\end{equation}
\begin{equation}
v =dr/dt = c(\frac{g_{00}}{g_{rr}})^{1/2} = c(1-r_g/r)
\label{15}
\end{equation}
[Note that if we defined (incorrectly!) $v$ as in Special Relativity,
$v=pc^2/E = c p_r/p_0$,
we would get a different behaviour of velocity: $v=c(1-r_g/r)^{-1}$, because
momentum increases both for the photon and for the stone.]

The time $t$ is often called the coordinate time, or world time, as it
is set by clocks infinitely far from the gravitating body. Therefore it
is proper to call $v$ the coordinate or world velocity. It is worth noting
that while the velocity $v$ changes with the potential, the
velocity of light in a locally inertial reference frame is always equal to
$c$.

The decrease of coordinate velocity $v$ discussed above leads to two types of
effects both of which are well known. One of them is the famous deflection of
light by the sun, predicted by Einstein and first observed during 1919 solar
eclipse. This observation was reported by newspapers worldwide and started
the cult of Einstein. The other effect is not so widely known. It is the
delay of radar echo from inner planets (Venus, Mercury), the measurement of
which was proposed \cite{6} and realized by I.I.Shapiro.
(For the description of the measurements see ref. \cite{3}, Section 8.7.)

It is appropriate to discuss
both effects (the echo and the deflection) in the
framework of geometrical optics (see ref. \cite{5}), as the wave-length
$\lambda$ is
negligible
compared to any other characteristic length. The refraction index $n$ is
given by:
\begin{equation}
n = \frac{c}{v} =(1-r_g /r)^{-1} =(1+r_g/r) \;\; .
\label{15a}
\end{equation}
Here and in what follows we consistently neglect terms of the order of
$(r_g/r)^2$: the gravitational field in the solar system is weak, and
therefore
\begin{equation}
n-1 = r_g/r \ll 1 \;\;.
\label{15''}
\end{equation}

\section{Radar echo}

Let us start with radar echo and consider for simplicity the echo from the
sun. (This is of course only a gedanken experiment). The duration of a two
way travel of a signal in this case is
\begin{equation}
t= 2\int\limits^{r_e}_{R_s} \frac{dr}{v(r)} =
\frac{2}{c}\int\limits^{r_e}_{R_s} n(r)dr =
\frac{2}{c}\int\limits^{r_e}_{R_s} (1+r_g/r)dr
\label{16}
\end{equation}
and the delay compared to the case $v=c$ is
\begin{equation}
\Delta t = 2\frac{r_g}{c}\ln\frac{r_e}{R_s}  \;\; ,
\label{17}
\end{equation}
where $r_e = 150 \cdot 10^6$ km is the distance between the sun and the
earth, while $R_s = 0.7 \cdot 10^6$ km is the radius of the sun.

Consider now the echo from a planet close to its superior conjunction (when
the planet is at the largest distance from the earth, so that the radar ray
almost grazes the sun). In that case, unlike the previous ones, we have a
non-radial trajectory with a non vanishing impact parameter $\rho$. Let us
denote by $z$ the coordinate along the straight line connecting the earth at
$(z_e, \rho)$ and the planet at $(z_p, \rho)$. (The axis $\rho$ is obviously
orthogonal to $z$.) Note that we can use straight lines even for non-radial
trajectories because the gravitational field in solar system is weak: $(n-1)
= r_g/r \ll 1$. The deflection of light (which is considered in the next
section) gives a negligible contribution of order $(n-1)^2$ to the echo delay
time. To calculate this delay time we use isotropic metric:
$$
\Delta t= \frac{2}{c}\int\limits^{z_p}_{z_e} dz(\frac{c}{v}-1) =
\frac{2}{c} \int\limits^{z_p}_{z_e}dz(r_g/r) =
$$
$$
2 \frac{r_g}{c} \int\limits^{z_p}_{z_e} \frac{dz}{\sqrt{z^2 +\rho^2}} =
2 \frac{r_g}{c}\left( \ln \frac{|z_p| +r_p}{\rho} +
\ln \frac{|z_e|+r_e}{\rho}\right) =
$$
\begin{equation}
= 2 \frac{r_g}{c} \ln \frac{4r_p r_e}{\rho^2} \;\; .
\label{18}
\end{equation}
The effect is maximal when $\rho \simeq R_s$.
The logarithmic dependence for
the echo from the sun and from a planet is the same. The
non-logarithmic terms ($\sim r_g/r_e$) are neglected here: we do not take
into account the retardation of clocks on the earth with respect to the
world time. Their inclusion would increase $\Delta t$ by $\sim$9\%. For
Mercury $r_p = 58\cdot 10^6$ km, hence $\Delta t \simeq 240\mu s$.

It is interesting to note that retardation must take place not only
for photons but also for any extreme relativistic particle if its mass $m$
and energy $E$ are such that
\begin{equation}
(mc^2)^2/2 E^2 \ll r_g/r_e
\label{19}
\end{equation}
For the sun - earth system $r_g/r_e \sim 2\cdot 10^{-8}$, while for electrons
from LEP I $(mc^2/E)^2 \sim 10^{-10}$. It is easy to see that even for low
energy solar and reactor neutrinos the inequality is also fulfilled.
Unfortunately echo experiments with electrons and neutrinos are not realistic
for many reasons.

\section{Deflection of light}

Consider a light ray connecting a star and the earth. In this case the
trajectory of light is bent, but we will continue to denote the variable
along this trajectory as $z$ (its end points $z_*$ and $z_e$, respectively)
and the impact parameter as $\rho$. Assume that we deal with a``thick pencil"
ray of diameter $d\rho$. It is easy to see from Eq. (\ref{15}) that the
inward side of the ray moves slower than the outward one.  (Here inward means
closer to the sun.) As a result the cross-section of the ray is tilted:  when
its inward edge propagates by $dz$ the outer edge propagates by
$dz[1+(\partial v/c\partial\rho)d\rho]$. Thus the angle of the tilt is
\begin{equation}
d\alpha = -dz(\partial v/c\partial\rho) = \frac{dz
\partial(r_g/r)}{\partial\rho} \;\; .
\label{20}
\end{equation}

Hence the direction of the ray would be inward deflected, the
differential deflection angle of the ray being given by Eq. (\ref{20}),
while the total deflection angle
\begin{equation}
\alpha = \int\limits^{z_e}_{z_*} dz
\frac{\partial(r_g/r)}{\partial\rho} =
\frac{\partial}{\partial\rho}\int\limits^{z_e}_{z_*}dz(r_g/r) \;\; .
\label{21}
\end{equation}

If we now substitute $z_*$ by $z_p$ which is not essential, because both of
them are much larger than $\rho$ and will not enter the result, we get from
Eq. (\ref{18}):
\begin{equation}
\alpha = \frac{\partial}{\partial\rho}\left(\frac{c\Delta t}{2}\right) =
-\frac{2r_g}{\rho} \;\; .
\label{22}
\end{equation}

As was stressed in ref. \cite{5}, the first part of Eq. (\ref{22}) is a
consequence not of General Relativity, but of geometrical optics.

For
the case of solar eclipse $\rho =R_s$ and
\begin{equation}
\alpha =-2 r_g/R_s = -1.75^{\prime\prime} \;\; .
\label{24}
\end{equation}

This is exactly the angle predicted by Einstein \cite{7} in 1915 in the
framework of General Relativity. He used the coordinate velocity $v$ given
by Eq.(\ref{15}). Note that in 1911 by implicitly assuming that $g_{rr}=1$ and,
hence, using $v=c(1-r_{g}/2r)$ Einstein derived an expression for the
angle in which coefficient 2 was missing \cite{8}.

\section{Conclusions}

Thus in a static gravitational field the frequency $\omega$ of the photon is
a constant; it is equal to the frequency at emission. The phenomenon of
gravitational redshift is explained by the increase of the energy difference
between levels in atoms or nuclei (in general, by the increase of the rate of
clocks) with the increase of their distance from the gravitating body. The
constant energy of photon appears as redshifted with respect to the
blueshifted energy difference of atomic levels.

The momentum of a photon $p$ and hence its wave-length $\lambda$ change with
the decrease of the distance from gravitating body: $p$ increases, $\lambda$
decreases. Hence the coordinate (world) velocity $v$ of the photon decreases.
This explains the two well established effects: the delay of radar echo from
planets and the deflection of star light by the sun.

\section{Acknowledgments}

I am grateful to A.D.Dolgov, V.L. Ginzburg, and K.G.Selivanov for insightful discussions. It
is a pity that the latter decidedly refused to coauthor this article. I also
would like to thank S.D.Drell, Ya.I.Granowski, J.D.Jackson and especially
V.L.Telegdi who read the draft of the paper for their constructive criticism.
This work has been partly supported by A. von Humboldt award and by RFBR grant
No.  00-15-96562.

\end{document}